\documentclass{article}

\usepackage[english]{babel}
\usepackage{amsmath,amssymb,amsfonts,amsthm}
\usepackage{hyperref}
\usepackage{graphicx}
\usepackage{url}
\usepackage{acronym}
\usepackage{subcaption}
\usepackage{colortbl}
\usepackage{xfrac}
\usepackage[inline]{enumitem}
\usepackage[numbers]{natbib}
\usepackage{textcomp}
\usepackage[htt]{hyphenat}
\usepackage{enumitem}

\definecolor{Gray}{gray}{0.80}
\definecolor{nfyellow}{rgb}{1.00, 0.75, 0.0}

\newcommand{\quotes}[1]{``.1"}

\begin{document}

\title{A comment to ``A General Theory of IR Evaluation Measures''}

\author{Fernando Giner\thanks{E.T.S.I. Inform\'{a}tica UNED, C/ Juan del Rosal, 16, 28040-Madrid, Spain (email: fginer3@gmail.com)}}

\maketitle

\begin{abstract}
The paper ``\emph{A General Theory of IR Evaluation Measures}'' develops a formal framework to determine whether IR evaluation measures are interval scales. This comment shows some limitations about its conclusions.
\end{abstract}

\section{Introduction}
Recently, the paper ``\emph{A General Theory of IR Evaluation Measures}''~\cite{FerranteEtAl2018b} has provided a formal framework, based on the \acf{RTM}~\citep{KrantzEtAl1971,LuceEtAl1990,SuppesEtAl1989}, for both set-based and rank-based IR evaluation measures as well as both  binary and multi-graded relevance, determining whether retrieval measures are interval scales.

However, some of its conclusions, such as ``\emph{Precision, Recall, and F-measure are interval scales, independently from the adopted order}'', only apply to some cases since there are several orderings, where these retrieval measures are not interval scales. The purpose of this comment is to point out some aspects that influence on the scale type of a retrieval measure when the \ac{RTM} is assumed. It is shown that the property of being an interval scale depends on the set where the retrieval measure is defined, more specifically, on the ordering relationship defined on this set\footnote{Note that this comment does not challenge other statements of the same authors about the scale type of retrieval measures, such as 'interval scales are equispaced mappings'~\cite{FerranteEtAl2021c}, which has been recently supported by providing a well documented discussion with the foundational works in the area of measurement~\cite{ferrante2022response}. This comment only points out that the arguments provided in~\citeauthor{FerranteEtAl2018b} are not generalisable.}.

On the other hand, I want to thank the great research line of work developed by the authors~\cite{FerranteEtAl2015b,FerranteEtAl2017b,FerranteEtAl2017,FerranteEtAl2018b,FerranteEtAl2019c,FerranteEtAl2021c}, which provides a better understanding of the theoretical foundations of IR evaluation. 

\section{Background on the RTM}
\label{sec:rtm}
The \ac{RTM} was initiated by~\citeauthor{von1887zahlen} and~\citeauthor{Holder1901} at the end of the 19th century. The \emph{opus magnum} of the \ac{RTM} can be found in the three volume work of Foundation of Measurement~\citep{KrantzEtAl1971,LuceEtAl1990,SuppesEtAl1989}; here, some of the main concepts are briefly summarized. 

The \ac{RTM} states that the numbers that are obtained through measurement represent empirical relationships. Two systems are considered: (i) the \acf{ERS}, which is a set of entities of the real world modelled with some relationships\slash operations; and (ii) the \acf{NRS}, which is commonly represented by the real numbers. Measurement consists on establishing a mapping between both systems, in such a way that the relationships between entities are matched by relationships between numbers. As there are different types of relationships, which can be represented on the \ac{NRS}, then there is a taxonomy on the representations, which leads to the notion of \emph{scale types}. 

Four main categories of scale types can be distinguished~\cite{Stevens1946} depending on the types of relationships\slash operations that are preserved: (i) a \emph{nominal scale} preserves every entity of the \ac{ERS} (one-to-one mappings); (ii) an \emph{ordinal scale} preserves an ordering relationship among entities (monotone increasing mappings); (iii) an \emph{interval scale} preserves a relationship among differences or intervals (linear positive mappings); and (iv) a \emph{ratio scale} preserves a relationship among ratios (similarity transformations). 

We formalise the concept of ordinal scale, as an example that will be referred later. Consider an \ac{ERS} endowed with an ordering relationship, $(X, \preceq)$, a real-mapping, $f$, is an ordinal scale if it preserves the ordering relationship on the \ac{ERS}, i.e., if it verifies: $x \preceq y$ iff $f(x) \leq f(y)$, $\forall x$, $y \in X$. Thus, the property of being an ordinal scale depends on the ordering relationship, $\preceq$, where the mapping is defined, and the same can be said for interval scales.

One of the main contributions of the \ac{RTM} are the \emph{representational theorems}, which determine the properties of the relationships\slash operations defined on the \ac{ERS}, such that they guarantee the existence and uniqueness of a mapping that preserves these relationships. For instance, in the particular case of interval scales, all the necessary and sufficient properties to obtain an interval scale are determined by a \emph{difference structure}, which is an \ac{ERS} endowed with an ordering relationship that verifies some specific properties~\citep[p.~59]{Rossi2014}.

\section{Brief Structure of the Paper}
\label{sec:structure}
The paper ``\emph{A General Theory of IR Evaluation Measures}''~\cite{FerranteEtAl2018b} considers measurement as a process of mapping lists of assessed documents (\ac{ERS}) to real numbers (\ac{NRS}), in such a way that the relationships defined on the ERS are represented or preserved as numerical properties. To determine whether retrieval measures are interval scales, a generic procedure is considered, which can be summarized with the following steps: 
\begin{enumerate}[label=(\arabic*)]
\item An ordering relationship among ranked lists of documents is defined. Thus, the \ac{ERS} is the set of lists of documents endowed with this relationship.
\item It is demonstrated that this \ac{ERS} is a difference structure.
\item One interval scale defined on this \ac{ERS} is identified\footnote{Note that the representation theorem of the \ac{RTM} assures that at least there exists one interval scale since the \ac{ERS} is a difference structure}.
\item Making use of the uniqueness of the representation theorem, it is checked whether a retrieval measure is an interval scale or not.
\end{enumerate}
The paper is structured and developed with the following four main sections: 
\begin{enumerate}[label=(\alph*)]
\item The set-based case is considered, \textbf{one} total order is defined by the authors, and the generic procedure is performed to deduce whether retrieval measures are interval scales on this \ac{ERS}.
\item The set-based case is considered again, but now, \textbf{one} partial order is defined, and the generic procedure is performed.
\item The rank-based case is considered, \textbf{one} total order is defined, and the generic procedure is performed to deduce whether retrieval measures are interval scales.
\item Finally, the rank-based case is considered again, \textbf{one} partial order is defined, and the generic procedure is performed.
\end{enumerate}
For each one of these four specific ordering relationships defined on the \ac{ERS}, conclusions are then drawn.

\section{Discussion}

\subsection{Main Issue}
Considering a retrieval measure, its property of being ordinal scale depends on the ordering defined on the \ac{ERS} since the retrieval measure has to preserve this ordering (see Section \ref{sec:rtm}). Thus, statements about the ordinal property of a retrieval measure, which do not specify the ordering on the \ac{ERS}, such as ``\emph{this retrieval measure is an ordinal scale.}'', have to be understood as considering all possible orderings on the \ac{ERS}, i.e., they should be interpreted as  ``\emph{this retrieval measure is an ordinal scale on all possible orderings on the \ac{ERS}}''. 

Otherwise, when a specific ordering on the \ac{ERS} is established or deduced from the context, then statements about the ordinal scale property of a retrieval measure should specify the considered ordering. In this case, statements of the form ``\emph{this retrieval measure is an ordinal scale.}'' are partially specified since they have to indicate the ordering on which the retrieval measure is defined. This statement should be expressed as ``\emph{this retrieval measure is an ordinal scale on} that \emph{empirical domain}'', where ``that'' is a specific \ac{ERS}.

These considerations about the ordinal property of a retrieval measure are also valid for the interval property since every interval scale is an ordinal scale. As noted in Section \ref{sec:structure},~\citeauthor{FerranteEtAl2018b} consider four scenarios: one total/partial order at the set and rank based cases, i.e., four specific orderings on the \ac{ERS} are considered. Thus, in this setting, it should not be made general statements of the form ``\emph{this retrieval measure is an interval scale.}'', neither it should be stated that the interval scale property of a retrieval measure is independent of the ordering on the \ac{ERS}.

However, some conclusions of the considered paper seem to express generic statements of the form: ``\emph{this retrieval measure is an interval scale.}''. In fact, these statements have been made clear in subsequent papers, such as in Ferrante et al.~\cite{FerranteEtAl2021c}: ``\emph{Recently, Ferrante et al. [40], [41] have theoretically shown that some of the most known and used IR measures, like Average Precision (AP) or Discounted Cumulative Gain (DCG), are not interval-scales.}''. 

The main issue about the setting of the considered paper consists in its lack of generalisation. When an ordering has been established on the empirical domain of a retrieval measure, then statements about its scale property should specify the \ac{ERS}, i.e., which are the assumptions on this set or its relationships and\slash or operators.

\subsection{Concerns at the Conclusions Section}
One of the conclusions about the set-based case on a total order claims that ``\emph{In the case of set-based \ac{IR} measures, binary relevance: Precision, Recall, and F-measure are interval scales, independently from the adopted order}''. However, as indicated in the previous subsection, the scale property of a retrieval measure is conditioned by the ordering on the \ac{ERS}. In fact, the generic procedure described in Section \ref{sec:structure} is not independent of the adopted order since it establishes a specific order at the step (1). 

Specifically, in the context of the quoted conclusion of the previous paragraph, i.e., in the case of set-based \ac{IR} measures with binary relevance, there are many orderings (some of them total orders) among lists of assessed documents where \ac{P}, \ac{R}, and F-measure are not interval scales. For instance, when the length of the retrieved lists is two ($N=2$), then it can be considered the following total order on the \ac{ERS}\footnote{Denoting by ``$1$'' a relevant document an by ``$0$'' a non-relevant one.}:
\[
\{0, 1\} \preceq \{0, 0\} \preceq \{1, 1\}. 
\]
However, \ac{P}, \ac{R}, and F-measure are not ordinal scales on this \ac{ERS} since \ac{P}$(\{0, 0\}) < $ \ac{P}$(\{0, 1\})$ (see definition of ordinal scale in Section \ref{sec:rtm}).

The main concern with the proposed procedure is its lack of generalisation since it only considers one ordering defined at the step (1). Once this ordering relationship has been established, then the interval scales deduced with this procedure, at the steps (3) and (4), are interval scales derived from this specific ordering. A retrieval measure may have different scale properties, depending on the empirical domain on which is defined. For instance, as it can be checked in the considered paper, the \ac{RBP} with $p=0.5$ is an interval scale when it is defined on the total ordering of the subsection 6.1, but not when it is defined on the partial ordering of the subsection 6.2. Therefore, there is a strong dependence between the scale type and the \ac{ERS} where the numerical mapping is defined.

Another statement made at the conclusion section is as follows: ``\emph{in the case of rank-based \ac{IR} measures, binary relevance: when using a total order, \ac{RBP} is an interval scale only if $p=\frac{1}{2}$ while all the other measures - namely AP, DCG, ERR, and RBP for other values of p - are not. When using a partial order, none of these measures is an interval scale.}''. The same comments can be made here, it is only considered one total order and one partial order, which are the orderings defined at the step (1), among the huge quantity of possible orderings. However, the conclusion claim is generic, and considers all possible total and partial orders (``\emph{when using \textbf{a} total order}'' or ``\emph{when using \textbf{a} partial order}''). Thus, the considered paper have only demonstrated that the previous retrieval measures are interval scales or not when they are defined on \textbf{one} specific ordering, but not that they are interval scales in general.

The distinction between the different ways of making statements about the scale properties of retrieval measures can be seen with the following two settings. On the one hand, in the foundational works of the \ac{RTM}~\citep{KrantzEtAl1971,LuceEtAl1990,SuppesEtAl1989}, the \ac{ERS} is initially endowed with a generic ordering relationship, i.e., no assumptions are adopted on this ordering. Thus, it can be considered that the \ac{ERS} is endowed with any possible ordering relationship. In this setting it is possible to make generic statements of the form ``\emph{this retrieval measure is an interval scale.}''. On the other hand, in~\citeauthor{FerranteEtAl2018b}'s paper, the \ac{ERS} is endowed with the ordering defined at step (1). Thus, the initial \ac{ERS} from which interval scales are deduced is an empirical domain endowed with a particular ordering. Therefore, the deduced interval scales, are interval scales defined on this specific \ac{ERS}. 

\subsection{Effect of these concerns}
Consider an \ac{IR} setting, where the \ac{ERS} is the set of possible system outputs rankings endowed with one of the ordering relationships of~\citeauthor{FerranteEtAl2018b}'s paper, $\preceq$. Pairs of system outputs rankings can be compared according to the users' information needs. If these comparisons between pairs of rankings do not hold, or hold only in some cases regarding $\preceq$, then it should be expected that the inferences drawn from the numerical values not to hold, or to hold only in some cases. As a consequence, incorrectly classifying the scale type of an IR evaluation measure leads to wrong conclusions about the obtained results.

Establishing an ordering on the \ac{ERS}, according to an intuitive and commonly agreeable relationship, as it has been done in the considered paper, limits the generalisation of the deduced interval scales.

\subsection{Other Concerns}
As a minor issue, the paper assumes that all rankings have a fixed number of documents, $N$, and conclusions about the scale properties of retrieval measures are based on this fact. However, different \ac{IR} systems can retrieve a different number of documents; thus, the analysis has a limited application. 

\section{Conclusions}
\label{sec:conclusion}
Guided by the goal ``\emph{to theoretically ground our evaluation methodology [...] in order to aim for more robust and generalizable inferences}''~\cite{ferrante2022response}, this comment has shown the difficulty of making generic statements of the form ``this retrieval measure is an interval scale.'' since, when the \ac{RTM} is assumed, the scale properties of retrieval measures depend on the \ac{ERS}. 

We think that this issue could be addressed by considering all the possible empirical domains where a retrieval measure is defined, or by letting the empirical domain to be specified by the retrieval measure itself, which is an interesting topic for future works.

\acrodef{3G}[3G]{Third Generation Mobile System}
\acrodef{5S}[5S]{Streams, Structures, Spaces, Scenarios, Societies}
\acrodef{AA}[AA]{Active Agreements}
\acrodef{AAAI}[AAAI]{Association for the Advancement of Artificial Intelligence}
\acrodef{AAL}[AAL]{Annotation Abstraction Layer}
\acrodef{AAM}[AAM]{Automatic Annotation Manager}
\acrodef{AAP}[AAP]{Average Average Precision}
\acrodef{ACLIA}[ACLIA]{Advanced Cross-Lingual Information Access}
\acrodef{ACM}[ACM]{Association for Computing Machinery}
\acrodef{AD}[AD]{Active Disagreements}
\acrodef{ADSL}[ADSL]{Asymmetric Digital Subscriber Line}
\acrodef{ADUI}[ADUI]{ADministrator User Interface}
\acrodef{AI}[AI]{Artificial Intelligence}
\acrodef{AIP}[AIP]{Archival Information Package}
\acrodef{AJAX}[AJAX]{Asynchronous JavaScript Technology and \acs{XML}}
\acrodef{ALS}[ALS]{Amyotrophic Lateral Sclerosis}
\acrodef{ALSFRS-R}[ALSFRS-R]{ALS Functional Rating Scale Revisited}
\acrodef{ALU}[ALU]{Aritmetic-Logic Unit}
\acrodef{AMUSID}[AMUSID]{Adaptive MUSeological IDentity-service}
\acrodef{ANOVA}[ANOVA]{ANalysis Of VAriance}
\acrodef{ANSI}[ANSI]{American National Standards Institute}
\acrodef{AP}[AP]{Average Precision}
\acrodef{APC}[APC]{AP Correlation}
\acrodef{API}[API]{Application Program Interface}
\acrodef{AR}[AR]{Address Register}
\acrodef{AS}[AS]{Annotation Service}
\acrodef{ASAP}[ASAP]{Adaptable Software Architecture Performance}
\acrodef{ASI}[ASI]{Annotation Service Integrator}
\acrodef{ASL}[ASL]{Achieved Significance Level}
\acrodef{ASM}[ASM]{Annotation Storing Manager}
\acrodef{ASR}[ASR]{Automatic Speech Recognition}
\acrodef{ASUI}[ASUI]{ASsessor User Interface}
\acrodef{ATIM}[ATIM]{Annotation Textual Indexing Manager}
\acrodef{AUC}[AUC]{Area Under the ROC Curve}
\acrodef{AUI}[AUI]{Administrative User Interface}
\acrodef{AWARE}[AWARE]{Assessor-driven Weighted Averages for Retrieval Evaluation}
\acrodef{BANKS-I}[BANKS-I]{Browsing ANd Keyword Searching I}
\acrodef{BANKS-II}[BANKS-II]{Browsing ANd Keyword Searching II}
\acrodef{BH}[BH]{Benjamini-Hochberg}
\acrodef{bpref}[bpref]{Binary Preference}
\acrodef{BNF}[BNF]{Backus and Naur Form}
\acrodef{BPM}[BPM]{Bejeweled Player Model}
\acrodef{BRICKS}[BRICKS]{Building Resources for Integrated Cultural Knowledge Services}
\acrodef{CAN}[CAN]{Content Addressable Netword}
\acrodef{CAS}[CAS]{Content-And-Structure}
\acrodef{CBSD}[CBSD]{Component-Based Software Developlement}
\acrodef{CBSE}[CBSE]{Component-Based Software Engineering}
\acrodef{CB-SPE}[CB-SPE]{Component-Based \acs{SPE}}
\acrodef{CD}[CD]{Collaboration Diagram}
\acrodef{CD}[CD]{Compact Disk}
\acrodef{CDF}[CDF]{Cumulative Density Function}
\acrodef{CENL}[CENL]{Conference of European National Librarians}
\acrodef{CIDOC CRM}[CIDOC CRM]{CIDOC Conceptual Reference Model}
\acrodef{CIR}[CIR]{Current Instruction Register}
\acrodef{CIRCO}[CIRCO]{Coordinated Information Retrieval Components Orchestration}
\acrodef{CG}[CG]{Cumulated Gain}
\acrodef{CL}[CL]{Curriculum Learning}
\acrodef{CL-ESA}[CL-ESA]{Cross-Lingual Explicit Semantic Analysis}
\acrodef{CLAIRE}[CLAIRE]{Combinatorial visuaL Analytics system for Information Retrieval Evaluation}
\acrodef{CLEF1}[CLEF]{Cross-Language Evaluation Forum}
\acrodef{CLEF}[CLEF]{Conference and Labs of the Evaluation Forum}
\acrodef{CLIR}[CLIR]{Cross Language Information Retrieval}
\acrodef{CM}[CM]{Continuation Methods}
\acrodef{CMS}[CMS]{Content Management System}
\acrodef{CMT}[CMT]{Campaign Management Tool}
\acrodef{CNR}[CNR]{Italian National Council of Research}
\acrodef{CO}[CO]{Content-Only}
\acrodef{COD}[COD]{Code On Demand}
\acrodef{CODATA}[CODATA]{Committee on Data for Science and Technology}
\acrodef{COLLATE}[COLLATE]{Collaboratory for Annotation Indexing and Retrieval of Digitized Historical Archive Material}
\acrodef{CP}[CP]{Characteristic Pattern}
\acrodef{CPE}[CPE]{Control Processor Element}
\acrodef{CPU}[CPU]{Central Processing Unit}
\acrodef{CQL}[CQL]{Contextual Query Language}
\acrodef{CRP}[CRP]{Cumulated Relative Position}
\acrodef{CRUD}[CRUD]{Create--Read--Update--Delete}
\acrodef{CS}[CS]{Characteristic Structure}
\acrodef{CSM}[CSM]{Campaign Storing Manager}
\acrodef{CSS}[CSS]{Cascading Style Sheets}
\acrodef{CTR}[CTR]{Click-Through Rate}
\acrodef{CU}[CU]{Control Unit}
\acrodef{CUI}[CUI]{Client User Interface}
\acrodef{CV}[CV]{Cross-Validation}
\acrodef{DAFFODIL}[DAFFODIL]{Distributed Agents for User-Friendly Access of Digital Libraries}
\acrodef{DAO}[DAO]{Data Access Object}
\acrodef{DARE}[DARE]{Drawing Adequate REpresentations}
\acrodef{DARPA}[DARPA]{Defense Advanced Research Projects Agency}
\acrodef{DAS}[DAS]{Distributed Annotation System}
\acrodef{DB}[DB]{DataBase}
\acrodef{DBMS}[DBMS]{DataBase Management System}
\acrodef{DC}[DC]{Dublin Core}
\acrodef{DCG}[DCG]{Discounted Cumulated Gain}
\acrodef{DCMI}[DCMI]{Dublin Core Metadata Initiative}
\acrodef{DCV}[DCV]{Document Cut--off Value}
\acrodef{DD}[DD]{Deployment Diagram}
\acrodef{DDC}[DDC]{Dewey Decimal Classification}
\acrodef{DDS}[DDS]{Direct Data Structure}
\acrodef{DF}[DF]{Degrees of Freedom}
\acrodef{DFI}[DFI]{Divergence From Independence}
\acrodef{DFR}[DFR]{Divergence From Randomness}
\acrodef{DHT}[DHT]{Distributed Hash Table}
\acrodef{DI}[DI]{Digital Image}
\acrodef{DIKW}[DIKW]{Data, Information, Knowledge, Wisdom}
\acrodef{DIL}[DIL]{\acs{DIRECT} Integration Layer}
\acrodef{DiLAS}[DiLAS]{Digital Library Annotation Service}
\acrodef{DIRECT}[DIRECT]{Distributed Information Retrieval Evaluation Campaign Tool}
\acrodef{DKMS}[DKMS]{Data and Knowledge Management System}
\acrodef{DL}[DL]{Digital Library}
\acrodefplural{DL}[DL]{Digital Libraries}
\acrodef{DLMS}[DLMS]{Digital Library Management System}
\acrodef{DLOG}[DL]{Description Logics}
\acrodef{DLS}[DLS]{Digital Library System}
\acrodef{DLSS}[DLSS]{Digital Library Service System}
\acrodef{DM}[DM]{Data Mining}
\acrodef{DO}[DO]{Digital Object}
\acrodef{DOI}[DOI]{Digital Object Identifier}
\acrodef{DOM}[DOM]{Document Object Model}
\acrodef{DoMDL}[DoMDL]{Document Model for Digital Libraries}
\acrodef{DP}[DP]{Discriminative Power}
\acrodef{DPBF}[DPBF]{Dynamic Programming Best-First}
\acrodef{DR}[DR]{Data Register}
\acrodef{DRIVER}[DRIVER]{Digital Repository Infrastructure Vision for European Research}
\acrodef{DTD}[DTD]{Document Type Definition}
\acrodef{DVD}[DVD]{Digital Versatile Disk}
\acrodef{EAC-CPF}[EAC-CPF]{Encoded Archival Context for Corporate Bodies, Persons, and Families}
\acrodef{EAD}[EAD]{Encoded Archival Description}
\acrodef{EAN}[EAN]{International Article Number}
\acrodef{EBU}[EBU]{Expected Browsing Utility}
\acrodef{ECD}[ECD]{Enhanced Contenty Delivery}
\acrodef{ECDL}[ECDL]{European Conference on Research and Advanced Technology for Digital Libraries}
\acrodef{EDM}[EDM]{Europeana Data Model}
\acrodef{EG}[EG]{Execution Graph}
\acrodef{ELDA}[ELDA]{Evaluation and Language resources Distribution Agency}
\acrodef{ELRA}[ELRA]{European Language Resources Association}
\acrodef{EM}[EM]{Expectation Maximization}
\acrodef{EMMA}[EMMA]{Extensible MultiModal Annotation}
\acrodef{EPROM}[EPROM]{Erasable Programmable \acs{ROM}}
\acrodef{EQNM}[EQNM]{Extended Queueing Network Model}
\acrodef{ER}[ER]{Entity--Relationship}
\acrodef{ERR}[ERR]{Expected Reciprocal Rank}
\acrodef{ERS}[ERS]{Empirical Relational System}
\acrodef{ESA}[ESA]{Explicit Semantic Analysis}
\acrodef{ESL}[ESL]{Expected Search Length}
\acrodef{ETL}[ETL]{Extract-Transform-Load}
\acrodef{FAST}[FAST]{Flexible Annotation Service Tool}
\acrodef{FDR}[FDR]{False Discovery Rate}
\acrodef{FIFO}[FIFO]{First-In / First-Out}
\acrodef{FIRE}[FIRE]{Forum for Information Retrieval Evaluation}
\acrodef{FN}[FN]{False Negative}
\acrodef{FNR}[FNR]{False Negative Rate}
\acrodef{FOAF}[FOAF]{Friend of a Friend}
\acrodef{FORESEE}[FORESEE]{FOod REcommentation sErvER}
\acrodef{FP}[FP]{False Positive}
\acrodef{FPR}[FPR]{False Positive Rate}
\acrodef{FVC}[FVC]{Forced Vital Capacity}
\acrodef{FWER}[FWER]{Family-wise Error Rate}
\acrodef{GIF}[GIF]{Graphics Interchange Format}
\acrodef{GIR}[GIR]{Geografic Information Retrieval}
\acrodef{GAP}[GAP]{Graded Average Precision}
\acrodef{GLM}[GLM]{General Linear Model}
\acrodef{GLMM}[GLMM]{General Linear Mixed Model}
\acrodef{GMAP}[GMAP]{Geometric Mean Average Precision}
\acrodef{GoP}[GoP]{Grid of Points}
\acrodef{GPRS}[GPRS]{General Packet Radio Service}
\acrodef{gP}[gP]{Generalized Precision}
\acrodef{gR}[gR]{Generalized Recall}
\acrodef{gRBP}[gRBP]{Graded Rank-Biased Precision}
\acrodef{GT}[GT]{Generalizability Theory}
\acrodef{GTIN}[GTIN]{Global Trade Item Number}
\acrodef{GUI}[GUI]{Graphical User Interface}
\acrodef{GW}[GW]{Gateway}
\acrodef{HCI}[HCI]{Human Computer Interaction}
\acrodef{HDS}[HDS]{Hybrid Data Structure}
\acrodef{HIR}[HIR]{Hypertext Information Retrieval}
\acrodef{HIT}[HIT]{Human Intelligent Task}
\acrodef{HITS}[HITS]{Hyperlink-Induced Topic Search}
\acrodef{HMM}[HMM]{Hidden Markov Model}
\acrodef{HTML}[HTML]{HyperText Markup Language}
\acrodef{HTTP}[HTTP]{HyperText Transfer Protocol}
\acrodef{HSD}[HSD]{Honestly Significant Difference}
\acrodef{ICA}[ICA]{International Council on Archives}
\acrodef{ICSU}[ICSU]{International Council for Science}
\acrodef{IDF}[IDF]{Inverse Document Frequency}
\acrodef{iDPP}[iDPP@CLEF]{Intelligent Disease Progression Prediction at CLEF}
\acrodef{IDS}[IDS]{Inverse Data Structure}
\acrodef{IEEE}[IEEE]{Institute of Electrical and Electronics Engineers}
\acrodef{IEI}[IEI]{Istituto della Enciclopedia Italiana fondata da Giovanni Treccani}
\acrodef{IETF}[IETF]{Internet Engineering Task Force}
\acrodef{IIR}[IIR]{Interactive Information Retrieval}
\acrodef{IMS}[IMS]{Information Management System}
\acrodef{IMSPD}[IMS]{Information Management Systems Research Group}
\acrodef{indAP}[indAP]{Induced Average Precision}
\acrodef{infAP}[infAP]{Inferred Average Precision}
\acrodef{INEX}[INEX]{INitiative for the Evaluation of \acs{XML} Retrieval}
\acrodef{INS-M}[INS-M]{Inverse Set Data Model}
\acrodef{INTR}[INTR]{Interrupt Register}
\acrodef{IP}[IP]{Internet Protocol}
\acrodef{IPSA}[IPSA]{Imaginum Patavinae Scientiae Archivum}
\acrodef{IR}[IR]{Information Retrieval}
\acrodef{IRON}[IRON]{Information Retrieval ON}
\acrodef{IRON2}[IRON$^2$]{Information Retrieval On aNNotations}
\acrodef{IRON-SAT}[IRON-SAT]{\acs{IRON} - Statistical Analysis Tool}
\acrodef{IRS}[IRS]{Information Retrieval System}
\acrodef{ISAD(G)}[ISAD(G)]{International Standard for Archival Description (General)}
\acrodef{ISBN}[ISBN]{International Standard Book Number}
\acrodef{ISIS}[ISIS]{Interactive SImilarity Search}
\acrodef{ISJ}[ISJ]{Interactive Searching and Judging}
\acrodef{ISO}[ISO]{International Organization for Standardization}
\acrodef{ITU}[ITU]{International Telecommunication Union }
\acrodef{ITU-T}[ITU-T]{Telecommunication Standardization Sector of \acs{ITU}}
\acrodef{IV}[IV]{Information Visualization}
\acrodef{JAN}[JAN]{Japanese Article Number}
\acrodef{JDBC}[JDBC]{Java DataBase Connectivity}
\acrodef{JMB}[JMB]{Java--Matlab Bridge}
\acrodef{JPEG}[JPEG]{Joint Photographic Experts Group}
\acrodef{JSON}[JSON]{JavaScript Object Notation}
\acrodef{JSP}[JSP]{Java Server Pages}
\acrodef{JTE}[JTE]{Java-Treceval Engine}
\acrodef{KDE}[KDE]{Kernel Density Estimation}
\acrodef{KLD}[KLD]{Kullback-Leibler Divergence}
\acrodef{KLAPER}[KLAPER]{Kernel LAnguage for PErformance and Reliability analysis}
\acrodef{LAM}[LAM]{Libraries, Archives, and Museums}
\acrodef{LAM2}[LAM]{Logistic Average Misclassification}
\acrodef{LAN}[LAN]{Local Area Network}
\acrodef{LD}[LD]{Linked Data}
\acrodef{LEAF}[LEAF]{Linking and Exploring Authority Files}
\acrodef{LIDO}[LIDO]{Lightweight Information Describing Objects}
\acrodef{LIFO}[LIFO]{Last-In / First-Out}
\acrodef{LM}[LM]{Language Model}
\acrodef{LMT}[LMT]{Log Management Tool}
\acrodef{LOD}[LOD]{Linked Open Data}
\acrodef{LODE}[LODE]{Linking Open Descriptions of Events}
\acrodef{LpO}[LpO]{Leave-$p$-Out}
\acrodef{LRM}[LRM]{Local Relational Model}
\acrodef{LRU}[LRU]{Last Recently Used}
\acrodef{LS}[LS]{Lexical Signature}
\acrodef{LSM}[LSM]{Log Storing Manager}
\acrodef{LtR}[LtR]{Learning to Rank}
\acrodef{LUG}[LUG]{Lexical Unit Generator}
\acrodef{MA}[MA]{Mobile Agent}
\acrodef{MA}[MA]{Moving Average}
\acrodef{MACS}[MACS]{Multilingual ACcess to Subjects}
\acrodef{MADCOW}[MADCOW]{Multimedia Annotation of Digital Content Over the Web}
\acrodef{MAD}[MAD]{Mean Assessed Documents}
\acrodef{MADP}[MADP]{Mean Assessed Documents Precision}
\acrodef{MADS}[MADS]{Metadata Authority Description Standard}
\acrodef{MAP}[MAP]{Mean Average Precision}
\acrodef{MARC}[MARC]{Machine Readable Cataloging}
\acrodef{MATTERS}[MATTERS]{MATlab Toolkit for Evaluation of information Retrieval Systems}
\acrodef{MDA}[MDA]{Model Driven Architecture}
\acrodef{MDD}[MDD]{Model-Driven Development}
\acrodef{METS}[METS]{Metadata Encoding and Transmission Standard}
\acrodef{MIDI}[MIDI]{Musical Instrument Digital Interface}
\acrodef{MIME}[MIME]{Multipurpose Internet Mail Extensions}
\acrodef{ML}[ML]{Machine Learning}
\acrodef{MLE}[MLE]{Maximum Likelihood Estimation}
\acrodef{MLIA}[MLIA]{MultiLingual Information Access}
\acrodef{MM}[MM]{Machinery Model}
\acrodef{MMU}[MMU]{Memory Management Unit}
\acrodef{MODS}[MODS]{Metadata Object Description Schema}
\acrodef{MOF}[MOF]{Meta-Object Facility}
\acrodef{MP}[MP]{Markov Precision}
\acrodef{MPEG}[MPEG]{Motion Picture Experts Group}
\acrodef{MRD}[MRD]{Machine Readable Dictionary}
\acrodef{MRF}[MRF]{Markov Random Field}
\acrodef{MRR}[MRR]{Mean Reciprocal Rank}
\acrodef{MS}[MS]{Mean Squares}
\acrodef{MS2}[MS]{Multiple Sclerosis}
\acrodef{MSAC}[MSAC]{Multilingual Subject Access to Catalogues}
\acrodef{MSE}[MSE]{Mean Square Error}
\acrodef{MT}[MT]{Machine Translation}
\acrodef{MV}[MV]{Majority Vote}
\acrodef{MVC}[MVC]{Model-View-Controller}
\acrodef{NACSIS}[NACSIS]{NAtional Center for Science Information Systems}
\acrodef{NAP}[NAP]{Network processors Applications Profile}
\acrodef{NCP}[NCP]{Normalized Cumulative Precision}
\acrodef{nCG}[nCG]{Normalized Cumulated Gain}
\acrodef{nCRP}[nCRP]{Normalized Cumulated Relative Position}
\acrodef{nDCG}[nDCG]{Normalized Discounted Cumulated Gain}
\acrodef{nMCG}[nMCG]{Normalized Markov Cumulated Gain}
\acrodef{NESTOR}[NESTOR]{NEsted SeTs for Object hieRarchies}
\acrodef{NEXI}[NEXI]{Narrowed Extended XPath I}
\acrodef{NII}[NII]{National Institute of Informatics}
\acrodef{NISO}[NISO]{National Information Standards Organization}
\acrodef{NIST}[NIST]{National Institute of Standards and Technology}
\acrodef{NIV}[NIV]{Non-Invasive Ventilation}
\acrodef{NLP}[NLP]{Natural Language Processing}
\acrodef{NN}[NN]{Neural Network}
\acrodef{NP}[NP]{Network Processor}
\acrodef{NR}[NR]{Normalized Recall}
\acrodef{NRS}[NRS]{Numerical Relational System}
\acrodef{NS-M}[NS-M]{Nested Set Model}
\acrodef{NTCIR}[NTCIR]{NII Testbeds and Community for Information access Research}
\acrodef{OAI}[OAI]{Open Archives Initiative}
\acrodef{OAI-ORE}[OAI-ORE]{Open Archives Initiative Object Reuse and Exchange}
\acrodef{OAI-PMH}[OAI-PMH]{Open Archives Initiative Protocol for Metadata Harvesting}
\acrodef{OAIS}[OAIS]{Open Archival Information System}
\acrodef{OC}[OC]{Operation Code}
\acrodef{OCLC}[OCLC]{Online Computer Library Center}
\acrodef{OMG}[OMG]{Object Management Group}
\acrodef{OO}[OO]{Object Oriented}
\acrodef{OODB}[OODB]{Object-Oriented \acs{DB}}
\acrodef{OODBMS}[OODBMS]{Object-Oriented \acs{DBMS}}
\acrodef{OPAC}[OPAC]{Online Public Access Catalog}
\acrodef{OQL}[OQL]{Object Query Language}
\acrodef{ORP}[ORP]{Open Relevance Project}
\acrodef{OSIRIS}[OSIRIS]{Open Service Infrastructure for Reliable and Integrated process Support}
\acrodef{P}[P]{Precision}
\acrodef{P2P}[P2P]{Peer-To-Peer}
\acrodef{PA}[PA]{Passive Agreements}
\acrodef{PAMT}[PAMT]{Pool-Assessment Management Tool}
\acrodef{PASM}[PASM]{Pool-Assessment Storing Manager}
\acrodef{PC}[PC]{Program Counter}
\acrodef{PCP}[PCP]{Pre-Commercial Procurement}
\acrodef{PCR}[PCR]{Peripherical Command Register}
\acrodef{PD}[PD]{Passive Disagreements}
\acrodef{PDA}[PDA]{Personal Digital Assistant}
\acrodef{PDF}[PDF]{Probability Density Function}
\acrodef{PDR}[PDR]{Peripherical Data Register}
\acrodef{PEG}[PEG]{Percutaneous Endoscopic Gastrostomy}
\acrodef{PIR}[PIR]{Personalized Information Retrieval}
\acrodef{POI}[POI]{\acs{PURL}-based Object Identifier}
\acrodef{PoS}[PoS]{Part of Speech}
\acrodef{PAA}[PAA]{Proportion of Active Agreements}
\acrodef{PPA}[PPA]{Proportion of Passive Agreements}
\acrodef{PPE}[PPE]{Programmable Processing Engine}
\acrodef{PREFORMA}[PREFORMA]{PREservation FORMAts for culture information/e-archives}
\acrodef{PRIMAD}[PRIMAD]{Platform, Research goal, Implementation, Method, Actor, and Data}
\acrodef{PRIMAmob-UML}[PRIMAmob-UML]{mobile \acs{PRIMA-UML}}
\acrodef{PRIMA-UML}[PRIMA-UML]{PeRformance IncreMental vAlidation in \acs{UML}}
\acrodef{PROM}[PROM]{Programmable \acs{ROM}}
\acrodef{PROMISE}[PROMISE]{Participative Research labOratory  for Multimedia and Multilingual Information Systems Evaluation}
\acrodef{pSQL}[pSQL]{propagate \acs{SQL}}
\acrodef{PUI}[PUI]{Participant User Interface}
\acrodef{PURL}[PURL]{Persistent \acs{URL}}
\acrodef{QA}[QA]{Question Answering}
\acrodef{QE}[QE]{Query Expansion}
\acrodef{QoS-UML}[QoS-UML]{\acs{UML} Profile for QoS and Fault Tolerance}
\acrodef{QPA}[QPA]{Query Performance Analyzer}
\acrodef{QPP}[QPP]{Query Performance Prediction}
\acrodef{R}[R]{Recall}
\acrodef{RAM}[RAM]{Random Access Memory}
\acrodef{RAMM}[RAM]{Random Access Machine}
\acrodef{RBO}[RBO]{Rank-Biased Overlap}
\acrodef{RBP}[RBP]{Rank-Biased Precision}
\acrodef{RBTO}[RBTO]{Rank-Based Total Order}
\acrodef{RDBMS}[RDBMS]{Relational \acs{DBMS}}
\acrodef{RDF}[RDF]{Resource Description Framework}
\acrodef{REST}[REST]{REpresentational State Transfer}
\acrodef{REV}[REV]{Remote Evaluation}
\acrodef{RF}[RF]{Relevance Feedback}
\acrodef{RFC}[RFC]{Request for Comments}
\acrodef{RIA}[RIA]{Reliable Information Access}
\acrodef{RMSE}[RMSE]{Root Mean Square Error}
\acrodef{RMT}[RMT]{Run Management Tool}
\acrodef{ROM}[ROM]{Read Only Memory}
\acrodef{ROMIP}[ROMIP]{Russian Information Retrieval Evaluation Seminar}
\acrodef{RoMP}[RoMP]{Rankings of Measure Pairs}
\acrodef{RoS}[RoS]{Rankings of Systems}
\acrodef{RP}[RP]{Relative Position}
\acrodef{RR}[RR]{Reciprocal Rank}
\acrodef{RSM}[RSM]{Run Storing Manager}
\acrodef{RST}[RST]{Rhetorical Structure Theory}
\acrodef{RSV}[RSV]{Retrieval Status Value}
\acrodef{RTM}[RTM]{Representational Theory of Measurement}
\acrodef{RT-UML}[RT-UML]{\acs{UML} Profile for Schedulability, Performance and Time}
\acrodef{SA}[SA]{Software Architecture}
\acrodef{SAL}[SAL]{Storing Abstraction Layer}
\acrodef{SAMT}[SAMT]{Statistical Analysis Management Tool}
\acrodef{SAN}[SAN]{Sistema Archivistico Nazionale}
\acrodef{SASM}[SASM]{Statistical Analysis Storing Manager}
\acrodef{SBTO}[SBTO]{Set-Based Total Order}
\acrodef{SD}[SD]{Sequence Diagram}
\acrodef{SE}[SE]{Search Engine}
\acrodef{SEBD}[SEBD]{Convegno Nazionale su Sistemi Evoluti per Basi di Dati}
\acrodef{SEM}[SEM]{Standard Error of the Mean}
\acrodef{SERP}[SERP]{Search Engine Result Page}
\acrodef{SFT}[SFT]{Satisfaction--Frustration--Total}
\acrodef{SIL}[SIL]{Service Integration Layer}
\acrodef{SIP}[SIP]{Submission Information Package}
\acrodef{SKOS}[SKOS]{Simple Knowledge Organization System}
\acrodef{SM}[SM]{Software Model}
\acrodef{SME}[SME]{Statistics--Metrics-Experiments}
\acrodef{SMART}[SMART]{System for the Mechanical Analysis and Retrieval of Text}
\acrodef{SoA}[SoA]{Service-oriented Architectures}
\acrodef{SOA}[SOA]{Strength of Association}
\acrodef{SOAP}[SOAP]{Simple Object Access Protocol}
\acrodef{SOM}[SOM]{Self-Organizing Map}
\acrodef{SPARQL}[SPARQL]{Simple Protocol and RDF Query Language}
\acrodef{SPE}[SPE]{Software Performance Engineering}
\acrodef{SPINA}[SPINA]{Superimposed Peer Infrastructure for iNformation Access}
\acrodef{SPLIT}[SPLIT]{Stemming Program for Language Independent Tasks}
\acrodef{SPOOL}[SPOOL]{Simultaneous Peripheral Operations On Line}
\acrodef{SQL}[SQL]{Structured Query Language}
\acrodef{SR}[SR]{Sliding Ratio}
\acrodef{sRBP}[sRBP]{Session Rank Biased Precision}
\acrodef{SRU}[SRU]{Search/Retrieve via \acs{URL}}
\acrodef{SS}[SS]{Sum of Squares}
\acrodef{SSD}[s.s.d.]{statistically significantly different}
\acrodef{SSTF}[SSTF]{Shortest Seek Time First}
\acrodef{STAR}[STAR]{Steiner-Tree Approximation in Relationship graphs}
\acrodef{STON}[STON]{STemming ON}
\acrodef{SVM}[SVM]{Support Vector Machine}
\acrodef{TAC}[TAC]{Text Analysis Conference}
\acrodef{TBG}[TBG]{Time-Biased Gain}
\acrodef{TCP}[TCP]{Transmission Control Protocol}
\acrodef{TEL}[TEL]{The European Library}
\acrodef{TERRIER}[TERRIER]{TERabyte RetrIEveR}
\acrodef{TF}[TF]{Term Frequency}
\acrodef{TFR}[TFR]{True False Rate}
\acrodef{TLD}[TLD]{Top Level Domain}
\acrodef{TME}[TME]{Topics--Metrics-Experiments}
\acrodef{TN}[TN]{True Negative}
\acrodef{TO}[TO]{Transfer Object}
\acrodef{TP}[TP]{True Positve}
\acrodef{TPR}[TPR]{True Positive Rate}
\acrodef{TRAT}[TRAT]{Text Relevance Assessing Task}
\acrodef{TREC}[TREC]{Text REtrieval Conference}
\acrodef{TRECVID}[TRECVID]{TREC Video Retrieval Evaluation}
\acrodef{TTL}[TTL]{Time-To-Live}
\acrodef{UCD}[UCD]{Use Case Diagram}
\acrodef{UDC}[UDC]{Universal Decimal Classification}
\acrodef{uGAP}[uGAP]{User-oriented Graded Average Precision}
\acrodef{UI}[UI]{User Interface}
\acrodef{UML}[UML]{Unified Modeling Language}
\acrodef{UMT}[UMT]{User Management Tool}
\acrodef{UMTS}[UMTS]{Universal Mobile Telecommunication System}
\acrodef{UoM}[UoM]{Utility-oriented Measurement}
\acrodef{UPC}[UPC]{Universal Product Code}
\acrodef{URI}[URI]{Uniform Resource Identifier}
\acrodef{URL}[URL]{Uniform Resource Locator}
\acrodef{URN}[URN]{Uniform Resource Name}
\acrodef{USM}[USM]{User Storing Manager}
\acrodef{VA}[VA]{Visual Analytics}
\acrodef{VAIRE}[VAIR\"{E}]{Visual Analytics for Information Retrieval Evaluation}
\acrodef{VATE}[VATE$^2$]{Visual Analytics Tool for Experimental Evaluation}
\acrodef{VIRTUE}[VIRTUE]{Visual Information Retrieval Tool for Upfront Evaluation}
\acrodef{VD}[VD]{Virtual Document}
\acrodef{VDM}[VDM]{Visual Data Mining}
\acrodef{VIAF}[VIAF]{Virtual International Authority File}
\acrodef{VIM}[VIM]{International Vocabulary of Metrology}
\acrodef{VL}[VL]{Visual Language}
\acrodef{VoIP}[VoIP]{Voice over IP}
\acrodef{VS}[VS]{Visual Sentence}
\acrodef{W3C}[W3C]{World Wide Web Consortium}
\acrodef{WAN}[WAN]{Wide Area Network}
\acrodef{WHO}[WHO]{World Health Organization}
\acrodef{WLAN}[WLAN]{Wireless \acs{LAN}}
\acrodef{WP}[WP]{Work Package}
\acrodef{WS}[WS]{Web Services}
\acrodef{WSD}[WSD]{Word Sense Disambiguation}
\acrodef{WSDL}[WSDL]{Web Services Description Language}
\acrodef{WWW}[WWW]{World Wide Web}
\acrodef{XAI}[XAI]{eXplainable \acs{AI}}
\acrodef{XMI}[XMI]{\acs{XML} Metadata Interchange}
\acrodef{XML}[XML]{eXtensible Markup Language}
\acrodef{XPath}[XPath]{XML Path Language}
\acrodef{XSL}[XSL]{eXtensible Stylesheet Language}
\acrodef{XSL-FO}[XSL-FO]{\acs{XSL} Formatting Objects}
\acrodef{XSLT}[XSLT]{\acs{XSL} Transformations}
\acrodef{YAGO}[YAGO]{Yet Another Great Ontology}
\acrodef{YASS}[YASS]{Yet Another Suffix Stripper}


\begin{thebibliography}{27}
\providecommand{\natexlab}[1]{#1}
\providecommand{\url}[1]{\texttt{#1}}
\expandafter\ifx\csname urlstyle\endcsname\relax
  \providecommand{\doi}[1]{doi: #1}\else
  \providecommand{\doi}{doi: \begingroup \urlstyle{rm}\Url}\fi

\bibitem[Ferrante et~al.(2015)Ferrante, Ferro, and Maistro]{FerranteEtAl2015b}
M.~Ferrante, N.~Ferro, and M.~Maistro.
\newblock {Towards a Formal Framework for Utility-oriented Measurements of
  Retrieval Effectiveness}.
\newblock In J.~Allan, W.~B. Croft, A.~P. de~Vries, C.~Zhai, N.~Fuhr, and
  Y.~Zhang, editors, \emph{{Proc. 1st ACM SIGIR International Conference on the
  Theory of Information Retrieval (ICTIR 2015)}}, pages 21--30. {ACM Press, New
  York, USA}, 2015.
  
\bibitem[Ferrante et~al.(2017)Ferrante, Ferro, and Pontarollo]{FerranteEtAl2017b}
M.~Ferrante, N.~Ferro, and S.~Pontarollo.
\newblock {An Interval-Like Scale Property for IR Evaluation Measures}.
\newblock In \emph{{Proc. 8th International Workshop on Evaluating Information Access (EVIA 2017)}}, pages 10--15. {EVIA@ NTCIR, Tokio, Japan}, 2017.

\bibitem[Ferrante et~al.(2017)Ferrante, Ferro, and
  Pontarollo]{FerranteEtAl2017}
M.~Ferrante, N.~Ferro, and S.~Pontarollo.
\newblock {Are IR Evaluation Measures on an Interval Scale?}
\newblock In J.~Kamps, E.~Kanoulas, M.~de~Rijke, H.~Fang, and E.~Yilmaz,
  editors, \emph{{Proc. 3rd ACM SIGIR International Conference on the Theory of
  Information Retrieval (ICTIR 2017)}}, pages 67--74. {ACM Press, New York,
  USA}, 2017.

\bibitem[Ferrante et~al.(2019)Ferrante, Ferro, and
  Pontarollo]{FerranteEtAl2018b}
M.~Ferrante, N.~Ferro, and S.~Pontarollo.
\newblock {A General Theory of IR Evaluation Measures}.
\newblock \emph{{IEEE Transactions on Knowledge and Data Engineering (TKDE)}},
  31\penalty0 (3):\penalty0 409--422, March 2019.

\bibitem[Ferrante et~al.(2020)Ferrante, Ferro, and Losiouk]{FerranteEtAl2019c}
M.~Ferrante, N.~Ferro, and E.~Losiouk.
\newblock {How do interval scales help us with better understanding IR
  evaluation measures?}
\newblock \emph{{Information Retrieval Journal}}, 23\penalty0 (3):\penalty0
  289--317, June 2020.

\bibitem[Ferrante et~al.(2021)Ferrante, Ferro, and Fuhr]{FerranteEtAl2021c}
M.~Ferrante, N.~Ferro, and N.~Fuhr.
\newblock {Towards Meaningful Statements in IR Evaluation. Mapping Evaluation
  Measures to Interval Scales}.
\newblock \emph{{IEEE Access}}, 9:\penalty0 136182--136216, 2021.

\bibitem[Ferrante et~al.(2022)Ferrante, Ferro, and Fuhr]{ferrante2022response}
M.~Ferrante, N.~Ferro, and N.~Fuhr.
\newblock {Response to Moffat's Comment on ``Towards Meaningful Statements in IR Evaluation: Mapping Evaluation Measures to Interval Scales''}.
\newblock \emph{{arXiv}}, 10.48550/ARXIV.2212.11735, 2022.

\bibitem[H{\"o}lder(1901)]{Holder1901}
O.~H{\"o}lder.
\newblock {Die Axiome der Quantit{\"a}t und die Lehre vom Mass}.
\newblock \emph{{Berichte {\"u}ber die Verhandlungen der K{\"o}niglich
  S{\"a}chsischen Gesellschaft der Wissenschaften zu Leipzig,
  Mathematisch-Physikaliche Classe}}, 53:\penalty0 1--64, 1901.

\bibitem[Krantz et~al.(1971)Krantz, Luce, Suppes, and Tversky]{KrantzEtAl1971}
D.~H. Krantz, R.~D. Luce, P.~Suppes, and A.~Tversky.
\newblock \emph{{Foundations of Measurement. Additive and Polynomial
  Representations}}, volume~1.
\newblock {Academic Press, New York, USA}, 1971.

\bibitem[Luce et~al.(1990)Luce, Krantz, Suppes, and Tversky]{LuceEtAl1990}
R.~D. Luce, D.~H. Krantz, P.~Suppes, and A.~Tversky.
\newblock \emph{{Foundations of Measurement. Representation, Axiomatization,
  and Invariance}}, volume~3.
\newblock {Academic Press, New York, USA}, 1990.

\bibitem[Rossi(2014)]{Rossi2014}
G.~B. Rossi.
\newblock \emph{{Measurement and Probability. A Probabilistic Theory of
  Measurement with Applications}}.
\newblock {Springer-Verlag, New York, USA}, 2014.

\bibitem[Stevens(1946)]{Stevens1946}
S.~S. Stevens.
\newblock {On the Theory of Scales of Measurement}.
\newblock \emph{{Science, New Series}}, 103\penalty0 (2684):\penalty0 677--680,
  June 1946.

\bibitem[Suppes et~al.(1989)Suppes, Krantz, Luce, and Tversky]{SuppesEtAl1989}
P.~Suppes, D.~H. Krantz, R.~D. Luce, and A.~Tversky.
\newblock \emph{{Foundations of Measurement. Geometrical, Threshold, and
  Probabilistic Representations}}, volume~2.
\newblock {Academic Press, New York, USA}, 1989.

\bibitem[von Helmholtz et~al.(1887)]{von1887zahlen}
H.~von Helmholtz.
\newblock \emph{Z{\"a}hlen und Messen Erkenntnis--theoretisch betrachtet, Philosophische Aufs{\"a}tze Eduard Zeller gewidmet}
\newblock {Fuess}, Leipzig, 1887.

\end{thebibliography}
\end{document}